\documentclass[%
 aip,
 amsmath,amssymb,
 reprint,%
]{revtex4-1}

\usepackage{graphicx}
\usepackage{dcolumn}
\usepackage{bm}

\usepackage[utf8]{inputenc}
\usepackage[T1]{fontenc}
\usepackage{mathptmx}

\begin{document}

\preprint{AIP/123-QED}

\title{Critical review of the scientific and economic\\efficacy of Lithium Lorentz Force Accelerators}

\author{A. Angus}
 \email[]{aangus@mix.wvu.edu}
\affiliation{ 
Slipspace Industries Corp., Morgantown, WV 26507, USA
}%

\begin{abstract}
Electromagnetic propulsion technology has been thought to provide a potential form of future spacecraft propulsion for some time. In contrast to ion thrusters, which utilize the Coulomb force to accelerate positively charged species, electromagnetic propulsion systems utilize the Lorentz force to accelerate all species in a quasi-neutral state, providing significant technological benefits over ion thrusters. Several forms of electromagnetic propulsion have been researched and developed, such as the Variable Specific Impulse Magnetoplasma Rocket, pulsed inductive thrusters, and the electrodeless plasma thruster. One of the most promising forms of electromagnetic propulsion, however, has been the magnetoplasmadynamic thruster. Whereas other electromagnetic propulsion systems provide high specific impulse values but low thrust capabilities, magnetoplasmadynamic thrusters have demonstrated the potential for both high specific impulse values and high thrust densities. However, these thrusters are not without drawbacks and suffer from issues such as electrode erosion. A proposed subtype of these thrusters, known as the Lithium Lorentz Force Accelerator, has been shown to address some of these issues. As is demonstrated in this paper, mission duration is not notably improved by the use of Lithium Lorentz Force Accelerators except for mission distances beyond the capabilities of current propulsion technology. It is also shown that increasing the amount of batteries onboard a spacecraft does not necessarily decrease mission duration due to the specific power of current battery technology, which is on the order of $10^3$ W/kg, but that new developments in nuclear energy technology may allow these thrusters to become efficacious for missions for which current propulsion technology is insufficient.
\end{abstract}

\maketitle

\section{\label{sec:one}Introduction}

Substantial research on Lithium Lorentz Force Accelerators (LiLFAs) has been conducted during the past several decades, in particular by Edgar Y. Choueiri and several colleagues at the Princeton University Electric Propulsion and Plasma Dynamics Laboratory (EPPDyL), as well as the Moscow Aviation Institute (MAI). \cite{one,two,three,four,five,six,seven,eight,nine,ten,eleven,twelve,thirteen,fourteen,fifteen,sixteen,seventeen,eighteen,nineteen,twenty,twenty-one,twenty-two,twenty-three} As such, this paper will not serve as a detailed technical treatment of LiLFAs, but rather as a review of the current efficacy of their near-term implementation as a means of spacecraft propulsion.

\begin{figure}[htp]
\centering
\includegraphics[width=8.5cm]{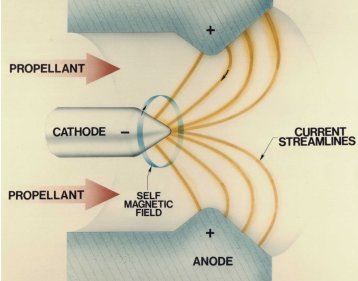}
\caption{\label{fig:one} Diagram of a magnetoplasmadynamic thruster. Reproduced from J. Gilland and G. Johnston, AIP Conf. Proc. \textbf{654}, 516 (2003), with the permission of AIP Publishing.}
\end{figure}

Magnetoplasmadynamic thrusters (MPDTs) were devised in the 1960s, before NASA had even reached the Moon.\cite{twenty-four} Most of this early research was highly theoretical, and the limits of energy technology during the era limited practical testing of these devices. MPDTs operate by using the Lorentz force to accelerate both the positively and negatively charged species of an ionized plasma through an exhaust nozzle at high velocities, which in some cases can exceed 100 km/s.\cite{twenty-five} This is achieved by utilizing an annular anode and cylindrical cathode, whereby a current is transmitted through the anode to the cathode, creating a magnetic field and a Lorentz force which has a strength that is based on the level of the applied current, $J$, as well as several other parameters (see Fig.~\ref{fig:one}).\cite{two}

The level of thrust generated by MPDTs is determined by several different factors that depend on whether the device is an applied-field (AF) configuration or a self-field (SF) configuration.\cite{two,eighteen} MPDTs inherently generate their own magnetic field; however, at low power levels this magnetic field is weak and the application of an additional magnetic field from an external source is required. These AF configurations are more near-term in nature and have been the focus of the majority of research conducted on MPDTs. At high power levels (>100 kW), however, the magnetic field that is induced by the thruster's discharge current is strong enough to produce a sufficient Lorentz force, and these SF configurations can operate at higher efficiencies than AF configurations.\cite{eighteen} A current drawback of SF configurations is that they require high power levels and are therefore limited by the current state-of-the-art in energy technology, as will be seen later.

Although MPDTs have shown a high level of promise as a form of future spacecraft propulsion, they suffer from electrode erosion issues due to the evaporation of the electrode surface. However, LiLFAs utilize a multi-channel hollow cathode system as opposed to the solid cathodes used in traditional MPDTs. This reduces the current density on the cathode surface, which reduces the energy density through the cathode material and the resulting issue of cathode erosion, which can be as high as 0.2 $\mu$g/C in solid cathode configurations.\cite{sixteen} As the name would suggest, LiLFAs use lithium vapor as propellant as opposed to the usual hydrogen. Lithium reduces the electrode temperature required to emit the discharge current, which helps further reduce electrode erosion while being more efficient, with specific impulse ($I_{sp}$) values of 1500-8000 s. This increased efficiency can exceed 40\% for applied-field Lithium Lorentz Force Accelerators (AF-LiLFAs), although these values are only present at power levels exceeding 500 kW.\cite{sixteen} This analysis will be restricted to LiLFA applications in the MW power range as these are more relevant to large-scale spacecraft. This also allows the study of self-field Lithium Lorentz Force Accelerators (SF-LiLFAs), which can provide higher efficiencies than AF-LiLFAs at sufficient power levels. LiLFAs also have thrust densities that are on the order of $10^5$ N/m\textsuperscript{2}, the highest for any form of electric propulsion.\cite{sixteen}

The main issue regarding the implementation of LiLFAs has historically been the lack of sufficient power sources in spacecraft, which has been well-documented over the past several decades.\cite{twelve,sixteen} In particular, there has been an absence of required steady-state power capabilities (in the MW range) as well as the required total energy capacity (on the order of $10^2$ MWh for sufficiently rapid missions to Mars). Sankaran \textit{et al}.\cite{twelve} conducted a survey of LiLFAs and other propulsion technologies in 2003 and reported these issues. However, a broad analysis of this technology had not been conducted in some time, and an additional analysis was required to determine any changes in the feasibility of LiLFAs. In order to conduct a broad analysis, astrodynamic simulations were performed in order to obtain a detailed understanding of the technical performance characteristics of LiLFAs, which are discussed in Sec.~\ref{sec:two}, followed by an analysis of the energy requirements and economic feasibility of these devices, which is discussed in Sec.~\ref{sec:three}. The analysis was then conducted again for a smaller spacecraft that used a small-scale AF-LiLFA, which is discussed in Sec.~\ref{sec:four}.

\section{\label{sec:two}Astrodynamic Models}

In order to fully understand the performance characteristics of LiLFAs, an analysis of their impact on mission duration in comparison to current propulsion technology was required. To begin this analysis, a theoretical spacecraft was introduced for these simulations. This spacecraft had a dry mass of $m_{dry}=60,000$ kg and a propellant mass of $m_{prop}=100,000$ kg, with these values being extrapolated from a broad range of historical spacecraft specifications. The spacecraft was powered by five SF-LiLFAs that each had an exhaust area of $A_e=2$ m\textsuperscript{2}, resulting in a thrust value of $T=200$ kN per thruster and a total thrust value of $T=1$ MN according to the thrust density of $10^5$ N/m\textsuperscript{2} established in Sec.~\ref{sec:one}. For MPDTs, $A_e$ is a function of both the anode radius, $r_a$, and the cathode radius, $r_c$. With this information, a transit time to Mars was calculated. Note that this analysis assumed the scenario of a robotic mission in which the spacecraft did not return to Earth. It is also important to note that hydrogen is thought to be a better propellant for large-scale MPDTs due to its suitability for high-power configurations as well as its higher exhaust velocities; however, it's low boiling point requires it to be cryogenically stored at extremely low temperatures to produce a high density.\cite{twenty-five} This presents technological challenges during long missions as it is difficult to prevent subcooled hydrogen from vaporizing, even with the use of complex equipment. However, even as a liquid, hydrogen has a substantially lower density than lithium, which can be stored as a solid at room temperature. These issues, combined with the cathode erosion issues mentioned above, mean that lithium is a better propellant for most scenarios.

\begin{figure}[htp]
\centering
\includegraphics[width=8.5cm]{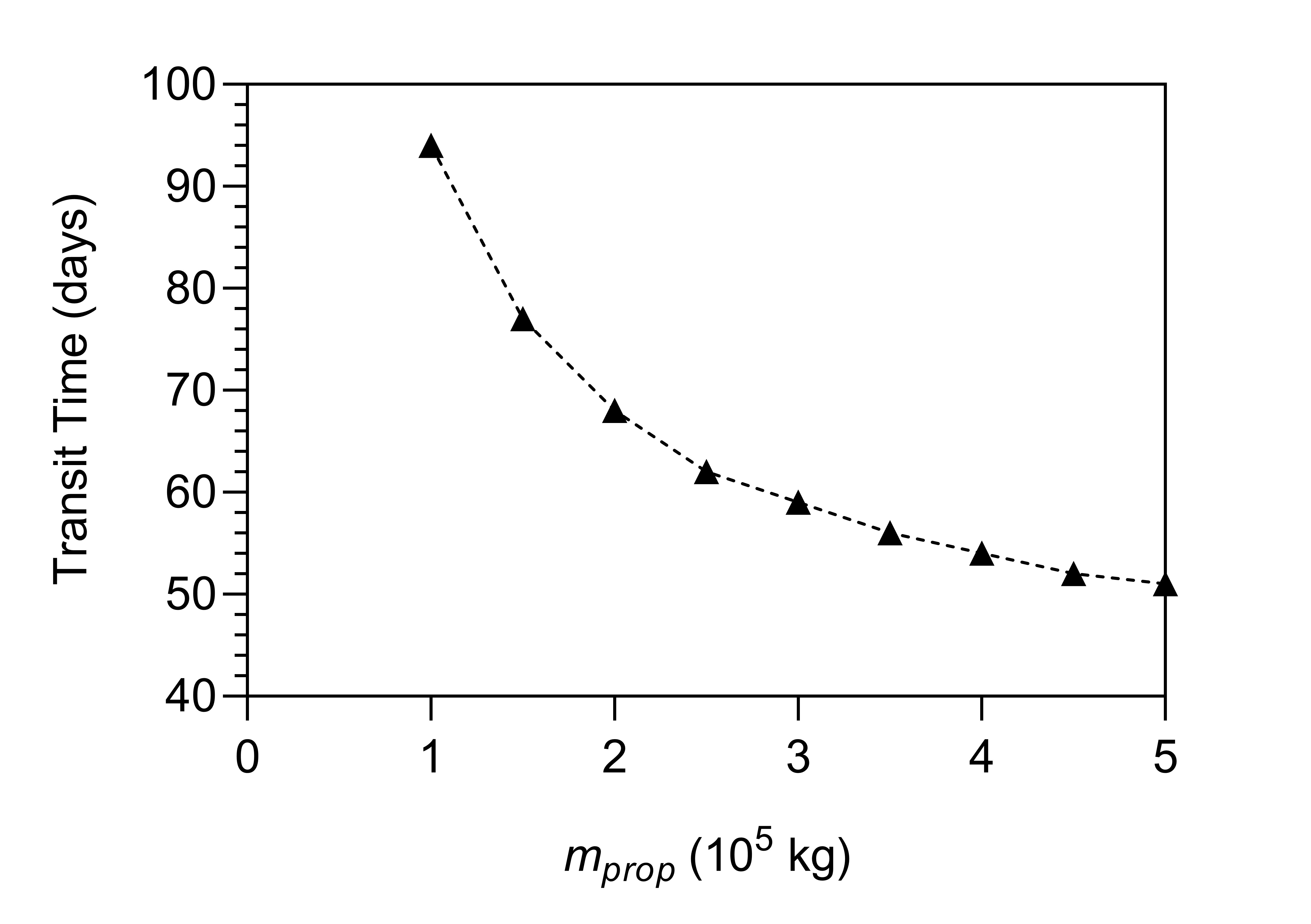}
\caption{\label{fig:two} Various transit times to Mars.}
\end{figure}

As these are SF-LiLFAs, the two components from which thrust is generated are
\begin{equation}
    T_{SF}=bJ^2\;,
    \label{eq:one}
\end{equation}
which is the SF component of thrust in which $J$ is the current through the electrodes and $b$ is a geometric scaling factor defined by
\begin{equation}
    b=\frac{\mu_0}{4\pi}
    \left[
    \ln
    \left(
    \frac{r_a}{r_c}
    \right)
    +\frac{3}{4}
    \right]\;;
    \label{eq:two}
\end{equation}
and
\begin{equation}
    T_{GD}=\dot{m}a_0+p_eA_e\;,
    \label{eq:three}
\end{equation}
where $a_0$ is the ion sound speed, $\dot{m}$ is the mass flow rate, and $p_e$ is the pressure at the exhaust exit.\cite{two} The parameter $\mu_0$ in Eq.~(\ref{eq:one}) is the permeability of free space. LiLFAs of this scale have yet to be tested, so to avoid complicating this discussion the total value of $\dot{m}$ for the LiLFAs the spacecraft used was determined to be 16.7 kg/s. Using Newton’s second law, this value was obtained from the expression $\dot{m}=(T-p_eA_e)/v_e$, using an exhaust velocity of $v_e=60$ km/s and neglecting the term $p_eA_e$ for simplicity (the thrust generated from this term would likely be negligible for large-scale SF-LiLFAs). The value of $v_e$ was obtained from an extrapolation of data in the literature and should provide a sufficient degree of accuracy for this analysis.\cite{two,five,eleven,twelve} To determine the transit time of the spacecraft, the General Mission Analysis Tool (GMAT), a high-fidelity mission analysis and trajectory optimization tool developed by NASA’s Goddard Space Flight Center, was used. GMAT allows for various types of mission simulations and allows users to determine unknown variables given a set of desired goals, as well as allowing users to optimize mission parameters. An electric thruster was used as the propulsion device using the parameters above and an $I_{sp}$ of 6104 s, which was obtained from the expression $I_{sp}=T/\dot{m}g_0$, where $g_0$ is the acceleration due to Earth’s gravity. This thruster was coupled with an electric propellant tank having the previously defined propellant mass of $m_{prop}=100,000$ kg. The departure date was initially chosen to be 5/5/2018 to allow a comparison of our spacecraft’s mission to that of the InSight robotic lander, the most recent mission to Mars. The departure date was then modified in order to converge on a solution for the new thrusting conditions, and additional code was written to optimize the solution. The trajectory was based on a B-Plane targeting of Mars, with the spacecraft decelerating before arrival to approximately 7000 m/s relative to Mars. GMAT automatically runs multiple iterations, varying the parameters until a solution is obtained within the desired margin of error, which in these simulations was within 0.1 km of the target destination. The simulation was run multiple times to ensure convergence validity and produced a total transit time of 94 days, approximately 46\% of the InSight lander’s transit time of 205 days. The simulation was performed for various values of $m_{prop}$, and the transit times for these values can be seen in Fig.~\ref{fig:two}.

The performance characteristics of large-scale SF configurations are therefore proven to be superior to that of current propulsion technology; however, as previously stated, the primary limiting factor regarding the implementation of these LiLFAs has been the lack of sufficient power in spacecraft. Although there have been significant advancements in spacecraft power over the past several decades, they still may not be enough to make large-scale SF-LiLFAs a feasible near-term form of spacecraft propulsion, as will be shown below.

\section{\label{sec:three}Analysis of Energy Technology}

Although the lack of sufficient power in spacecraft has long plagued the use of LiLFAs, improvements in energy technology by organizations in varying industries may have the consequence of changing the state-of-the-art in spacecraft propulsion. As such, an analysis of these advancements was conducted in order to better understand the efficacy of LiLFAs as their performance is directly impacted by the characteristics of their power source.

Of all the developments that have been made in a wide range of energy technologies during that past several years, some of particular relevance are the substantial advancements that have been made regarding lithium-ion (Li-ion) batteries. Nuclear energy has long represented the most promising option as a form of power for electric propulsion technology, especially for long missions that exceed the operating lifetime of battery technology. However, advancements in the specific power and specific energy of batteries suggests that this may no longer be the case. Companies such as Tesla, Inc. have made great improvements in the energy capacity and peak power delivery of Li-ion batteries for use in their electric vehicles (EVs). The United States Department of Energy (DOE) is also funding research in an effort to develop Li-ion batteries with a specific energy of $E_{sp}=500$ Wh/kg.\cite{twenty-six} A particularly promising form of these batteries is the new solid-state battery. Michael A. Zimmerman, a researcher at Tufts University, developed a solid polymer electrolyte to replace the liquid electrolyte in Li-ion batteries, and founded the company Ionic Materials, Inc. to research and develop this technology.\cite{twenty-seven} Additionally, the company Solid Power, Inc. is developing similar all-solid-state batteries (ASSBs), which replace the liquid electrolyte and plastic separators in Li-ion batteries. This provides more stability across a broad temperature range and allows for more efficient packaging, effectively increasing the energy density of the batteries. The company reports that their batteries will have an $E_{sp}$ of 320 to 700 Wh/kg, an energy density ($u$) of 700 to 1100 kWh/m\textsuperscript{3}, and a specific power ($P_{sp}$) exceeding 1 kW/kg (this data is according to the Solid Power, Inc. website as of the submission of this paper). Although other research is being conducted in the field of energy technology, particularly with respect to nuclear energy, battery technology has provided some of the most promising level of progress in recent years. Additionally, ASSBs currently seem to provide the highest level of performance in terms of specific power and specific energy. As such, these batteries were used to analyze the power and energy requirements of the simulated mission to Mars.

The approximate thrust to power ratio (also known as thrust power) of 25 N/MW for LiLFAs has been established for some time.\cite{eighteen} Using this ratio, the power requirement of the LiLFAs was determined to be $P=40$ GW. Then, using the total operating time of the LiLFA during the simulation discussed in Sec.~\ref{sec:two}, as well as the upper limits of the ASSBs under development by Solid Power, Inc., it was determined that we would need $8.08\times10^7$ kg (80,800 t) and $5.14\times10^4$ m\textsuperscript{3} of ASSBs to meet the energy requirement, or $4.00\times10^7$ kg (40,000 t) and $2.55\times10^4$ m\textsuperscript{3} of ASSBs to meet the power requirement. This represents substantial technological and unrealistic cost requirements, especially considering that this analysis only accounts for a one-way robotic mission to Mars. A manned mission would require additional propellant for the return trip, making these requirements even higher. Note that, in this case, it can be seen that the limiting factor is the total energy requirement.

\section{\label{sec:four}Analysis of Applied-Field Configurations}

\begin{figure}[htp]
\centering
\includegraphics[width=8.5cm]{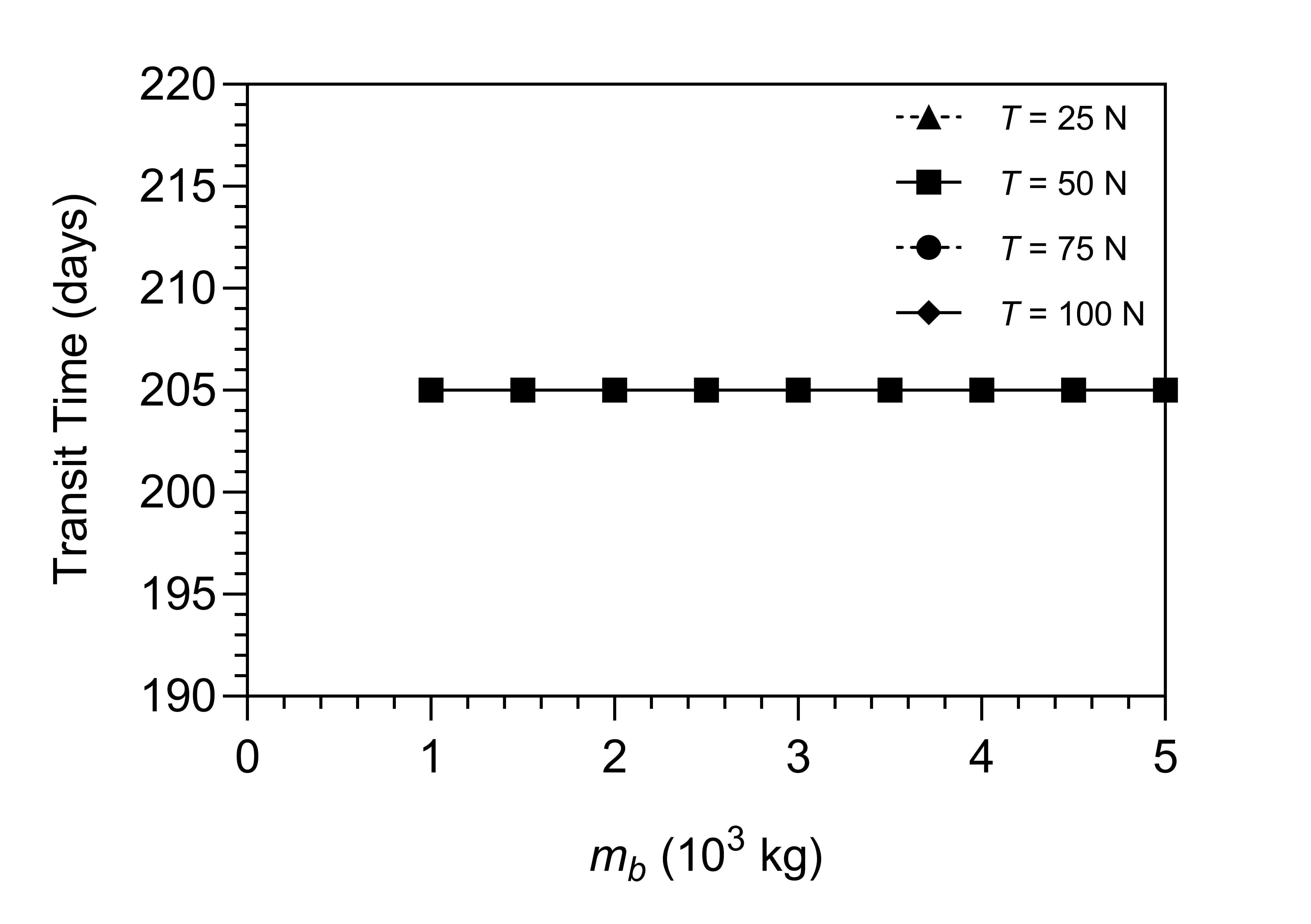}
\caption{\label{fig:three} Transit times to Mars for various values of $m_b$ and $T$.}
\end{figure}

The calculations made in Sec.~\ref{sec:two} and \ref{sec:three} were repeated with a smaller spacecraft utilizing a single small-scale AF-LiLFA. For this theoretical spacecraft, a bottom-up approach of beginning with known components was used as opposed to the top-down approach of beginning with known performance specifications as was used in Sec.~\ref{sec:two}, the reason for which will be explained later in this section. This spacecraft was designed to send a small rover to Mars as opposed to humans or large equipment and was assumed to serve as both an aeroshell and an integration structure for the LiLFA. A dry mass of $m_{dry}=2000$ kg was used, which included the rover as well as 1000 kg of ASSBs. Using the upper limits of the ASSBs, the spacecraft's power output and total energy capacity were determined to be $P=1$ MW and $E=700$ kWh, respectively. Using the previously established thrust to power ratio of 25 N/MW, the thrust output of the LiLFA was determined to be $T=25$ N, which corresponds to an exhaust area of $A_e=2.5$ cm\textsuperscript{2}. For reference, the AF thrust component is defined by
\begin{equation}
    T_{AF}=kJB_Ar_a\;,
    \label{eq:four}
\end{equation}
where $k$ is a scaling constant and $B_A$ is the applied magnetic field.\cite{two} However, the thrust to power ratio was used for simplicity. The value of $\dot{m}$ was again determined from the expression $\dot{m}=(T-p_eA_e)/v_e$. A lower exhaust velocity of $v_e=40$ km/s was chosen for this simulation as this value is in the range of reported exhaust velocities for AF configurations.\cite{two,five,eleven,twelve} Neglecting the term $p_eA_e$ again, the value of $\dot{m}=0.625$ g/s was obtained, which corresponds to a specific impulse of $I_{sp}=4077$ s. The LiLFA was limited to 2520 seconds of operation based on the total energy capacity of the ASSBs, which corresponded to a propellant mass of $m_{prop}=1.58$ kg. The GMAT simulation was performed again with these new parameters, producing a total transit time of 204.8 days, nearly identical to the InSight lander’s transit time. To further investigate this result, the simulation was performed for various masses of ASSBs. Transit times for various values of $m_b$ (mass of batteries) and T can be seen in Fig.~\ref{fig:three}.
It can be seen that the curve is relatively flat due to the particular relationship between this set of parameters, with each value differing by less than $10^{-2}$ days. This is due to the relatively negligible effect that the propulsion system had on the overall trajectory, with the velocity change ($\Delta v$) that was created by the propulsion system being small compared to the initial velocity gained from the Earth gravity assist. This lack of impact is a result of the low specific power of the ASSBs, which prevented the propulsion system from providing sufficient thrusting capabilities to create a large $\Delta v$. Further investigation lead to the conclusion that the specific power of the power source onboard a spacecraft has a substantial impact on the effectiveness of LiLFAs. This resulted in the derivation of a new metric to assess the performance of electric propulsion designs, which is expressed as
\begin{equation}
    a_{sp}=\frac{P_TP_{sp}}{m}\;,
    \label{eq:five}
\end{equation}
where $a_{sp}$ is the specific acceleration of the electric propulsion design, $P_T$ is the thrust power, and $m$ is the total mass of the spacecraft. Specific acceleration is the increase in acceleration per unit increase in the mass of the power system for a given spacecraft mass and is in units of m/(s\textsuperscript{2}$\cdot$kg). If we use the values for the AF-LiLFA discussed above, $a_{sp}$ becomes $1.25\times10^{-5}$ m/s\textsuperscript{2}$\cdot$kg. This low value explains the lack of influence the LiLFA exhibited during the simulations with the AF configuration. Note that the low value of $P_{sp}$ for the ASSBs also means that the top-down approach used in Sec.~\ref{sec:two} is flawed and that adding the $8.08\times10^7$ kg of batteries would increase the transit time to a similar value of 204.7 days. Therefore, the mass of the required power system must be known beforehand in order to accurately calculate the dry mass of the spacecraft when the former accounts for a large portion of the latter. Although these simulations neglected the small portion of power required to operate the propellant feeding system and the applied magnetic field (in the case of the AF-LiLFA), this power is not necessarily negligible. Also note that Eq.~(\ref{eq:five}) can be used to determine the specific acceleration of any electric propulsion configuration given the dry mass of the spacecraft.

As it appears that battery technology is still insufficient for use with large-scale LiLFAs, the next logical candidate would be nuclear energy. Although governments have historically been prohibited from using nuclear power in space due to various legislation, the political climate regarding small-scale nuclear power systems has warmed in recent years and governments are now beginning to research this technology. One device of particular interest is the Kilopower Reactor Using Stirling Technology (KRUSTY), a small fission reactor that is being researched by NASA and the DOE’s National Nuclear Security Administration. As the name would suggest, KRUSTY reactors use the Stirling cycle to generate electrical energy from the fission of uranium-235. The reactors are intended to be produced in four different sizes and will produce 1-10 kW, depending on the variant. They are also designed to be intrinsically safe and have a number of mechanisms to help reduce the risk of a nuclear meltdown, including a passive cooling system. Although the 10 kW variant’s mass of 1500 kg results in a specific power of $P_{sp}=6.67$ W/kg, which is substantially lower than the specific power of the ASSBs, its estimated 12-15 year lifespan results in an effectively unlimited energy capacity during the course of an interplanetary mission, meaning that the duration of acceleration is limited only by the mass of propellant. This can be especially beneficial for robotic missions that explore the far reaches of the solar system. Additionally, NASA reports that four of the 10 kW variants are sufficient to provide In-Situ Resource Utilization (ISRU) on Mars by separating and cryogenically storing oxygen from the Martian atmosphere for use as propellant, and also reports that the same reactors are sufficient to support a crew of 4-6 astronauts after the ISRU phase is complete.\cite{twenty-eight} This represents promising progress regarding the feasibility of a manned mission to Mars, and further advancements of these reactors will improve their feasibility as form of power for LiLFAs. The low specific power of early space-based nuclear reactors will present issues for shorter missions as the low acceleration capabilities provided by these devices will lead to the selection of other forms of propulsion; however, as the specific power of these devices improves, they will become feasible for a broader range of missions.

It would appear from this analysis that, from a technical standpoint, LiLFAs still represent one of the most promising options for future spacecraft propulsion as indicated by the potentially short transit times given a power source with sufficient performance characteristics. However, substantial advancements in energy technology must be made in order for these devices to become better than current propulsion technologies, especially in the case of large-scale SF-LiLFAs. In particular, the specific power of spacecraft power systems must improve drastically. Battery technology is yet to be sufficient for powering these devices, and other power options, such as the KRUSTY reactor, may represent a better technological investment. Note that the specific power of these reactors will need to increase substantially even for them to match that of current battery technology. A similar device known as the Radioisotope Thermoelectric Generator (RTG), which produces energy by capturing the heat released by a radioactive material, is typically used on missions to deep space, but the most recent variant used on the Curiosity rover has a specific power value of $P_{sp}=2.8$ W/kg, even lower than that of the KRUSTY reactor. Solar panels are another commonly proposed power source, and new advancements in solar panel technology could allow for thin-film solar panels with a specific power value as high as $P_{sp}=4.3$ kW/kg. However, the power output of solar panels decreases with increasing distance from the Sun, making them ineffective for mission to deep space. Fusion energy has also been the subject of extensive research during the past decade, and the successful development of fusion reactors could provide a substantial breakthrough for the field of electric propulsion; however, even if such devices are created, early variants will likely be far too large for spacecraft applications. Regardless of the form, if organizations such as NASA seek to use LiLFAs to conduct rapid exploration of the solar system, they must invest more heavily in the supplemental energy technology required for these devices to be effective.

\section{\label{sec:five}Conclusion}

As can be seen from the results in the previous sections, the current state-of-the-art in energy technology is still insufficient for the implementation of large-scale SF-LiLFAs by a considerable degree. Although significant advancements have been made over the past several years, and although research and development by organizations in various industries shows promising signs of continued advancement, an alternate form of energy may need to be considered for the successful implementation of these types of LiLFAs. The most obvious candidate would be nuclear energy. However, governments have historically been hesitant to use large-scale forms of nuclear energy in space due to political reasons, and as a result, current space-based nuclear power technology is small in scale and does not meet the requirements for large-scale LiLFAs. Until large-scale nuclear power systems are developed for space applications, or until the required advancements have been made in battery technology (which likely will not be for several decades), large-scale SF-LiLFAs will remain a long-term form of spacecraft propulsion, and manned missions will need to utilize other forms of propulsion. It is important to note that small-scale AF-LiLFAs still show promise as a near-term form of propulsion for small spacecraft, particularly for unmanned missions, provided that desired acceleration capabilities can be achieved.

Further investigation will provide more insight into the near-term readiness of this technology when coupled with small-scale nuclear reactors. Historically, organizations such as NASA have been required to balance competing interests and limited resources. However, the technological advancements that LiLFAs present may be significant, ultimately meaning that the perceived urgency of this technology is dependent upon the organization's goals. Regardless, as Princeton University’s EPPDyL and other organizations continue to make progress with this technology, a better understanding of LiLFAs will be gained and the state-of-the-art in the field will continue to advance.

\section*{Acknowledgments}

The author would like to thank Joseph Dygert, Dr. Patrick Browning, and Dr. Christopher Griffin of the Department of Mechanical and Aerospace Engineering at West Virginia University for their valuable advice, as well as the NASA Goddard Space Flight Center for providing GMAT.

\end{document}